\documentclass[11pt,twoside,titlepage,fleqn]{cslarticle}
\cslreportnumber{SRI-CSL Technical Report}
\date{November 2017, revised May 2022}
\acknowledge{John Rushby's research was partially supported by NASA
Contract NNL16AA06B, Task Order NNL16AA96T, under a subcontract to Honeywell.}
\usepackage{txfonts}
\usepackage[bookmarks=true,hyperfigures=true,colorlinks=true,linkcolor=blue,citecolor=blue,backref=page,pagebackref=false,pdfpagemode=fullscreen,plainpages=false,pdfpagelabels]{hyperref}
\usepackage[all]{hypcap}
\usepackage[normalem]{ulem} 
\usepackage{caption,cite,url,latexsym,alltt}
\usepackage[utf8]{inputenc}
\DeclareUnicodeCharacter{25A1}{\ensuremath\Box}
\DeclareUnicodeCharacter{25C7}{\ensuremath\Diamond}
\DeclareSymbolFont{symbolsC}{U}{txsyc}{m}{n}
\DeclareMathSymbol{\strictif}{\mathrel}{symbolsC}{74}
\def\smaller{\ifx\@currsize\Huge \protect\huge \else
              \ifx\@currsize\huge \protect\LARGE \else
               \ifx\@currsize\LARGE \protect\Large \else
                \ifx\@currsize\Large \protect\large \else
                 \ifx\@currsize\large \protect\normalsize \else
                  \ifx\@currsize\normalsize \protect\small \else
                   \ifx\@currsize\small \protect\footnotesize \else
                    \ifx\@currsize\footnotesize \protect\scriptsize \else
                       \protect\tiny
                    \fi\fi\fi\fi\fi\fi\fi\fi}
\def\larger{\ifx\@currsize\tiny \protect\scriptsize \else
              \ifx\@currsize\scriptsize \protect\footnotesize \else
               \ifx\@currsize\footnotesize \protect\small \else
                \ifx\@currsize\small \protect\normalsize \else
                 \ifx\@currsize\normalsize \protect\large \else
                  \ifx\@currsize\large \protect\Large \else
                   \ifx\@currsize\Large \protect\LARGE \else
                    \ifx\@currsize\LARGE \protect\huge \else
                       \protect\Huge
                    \fi\fi\fi\fi\fi\fi\fi\fi}
\def\BigLaTeX{{\rm L\kern-.36em\raise.3ex\hbox{\smaller\smaller A}\kern-.15em
    T\kern-.1667em\lower.7ex\hbox{E}\kern-.125emX}}
\def\BoldLaTeX{{\bf L\kern-.36em\raise.3ex\hbox{\smaller\smaller\bf A}\kern-.15em
    T\kern-.1667em\lower.7ex\hbox{E}\kern-.125emX}}
\def\BibTeX{{\rm B\kern-.05em{\sc i\kern-.025em b}\kern-.08em
    T\kern-.1667em\lower.7ex\hbox{E}\kern-.125emX}}

\newlength{\hsbw}
\newenvironment{session}{\begin{flushleft}
 \setlength{\hsbw}{\linewidth}
 \addtolength{\hsbw}{-\arrayrulewidth}
 \addtolength{\hsbw}{-\tabcolsep}
 \begin{tabular}{@{}|c@{}|@{}}\hline 
 \begin{minipage}[b]{\hsbw}
 \begingroup\sessionsize\vspace*{1.2ex}\begin{alltt}}{\end{alltt}\endgroup\end{minipage}\\ \hline
 \end{tabular}
 \end{flushleft}}
\def\extrawidth{0.5in}

\newenvironment{sessionlab}[1]{\begin{flushleft}
 \setlength{\hsbw}{\linewidth}
 \addtolength{\hsbw}{-\arrayrulewidth}
 \addtolength{\hsbw}{-\tabcolsep}
 \begin{tabular}{@{}|c@{}|@{}}\hline 
 \begin{minipage}[b]{\hsbw}
 \vspace*{-.5pt}
 \begin{flushright}
 \rule{0.01in}{.15in}\rule{0.9in}{0.01in}\hspace{-0.95in}
 \raisebox{0.04in}{\makebox[0.9in][c]{\footnotesize #1}}
 \end{flushright}
 \vspace*{-.47in}
 \begingroup\small\vspace*{1.0ex}\begin{alltt}}{\end{alltt}\endgroup\end{minipage}\\ \hline 
 \end{tabular}
 \end{flushleft}}
\newcounter{sessioncount}
\newenvironment{session*}{\begin{flushleft}
 \refstepcounter{sessioncount}
 \setlength{\hsbw}{\linewidth}
 \addtolength{\hsbw}{-\arrayrulewidth}
 \addtolength{\hsbw}{-\tabcolsep}
 \begin{tabular}{@{}|c@{}|@{}}\hline 
 \begin{minipage}[b]{\hsbw}
 \vspace*{-.5pt}
 \begin{flushright}
 \rule{0.01in}{.15in}\rule{0.3in}{0.01in}\hspace{-0.35in}
 \raisebox{0.04in}{\makebox[0.3in][c]{\footnotesize \thesessioncount}}
 \end{flushright}
 \vspace*{-.57in}
 \begingroup\small\vspace*{1.0ex}\begin{alltt}}{\end{alltt}\endgroup\end{minipage}\\ \hline 
 \end{tabular}
 \end{flushleft}}
\def\sessionsize{\small}
\def\smallsessionsize{\small}

\newcommand{\exmemo}[1]{}

\newcommand{\comment}[1]{}

\newcommand{\exfootnote}[1]{}
\newlength{\sblen}
\newlength{\overhang}

\def\SetFigFont#1#2#3{\rm}

\newcommand{\excite}[1]{}

\def\srilogo{\includegraphics[height=18mm]{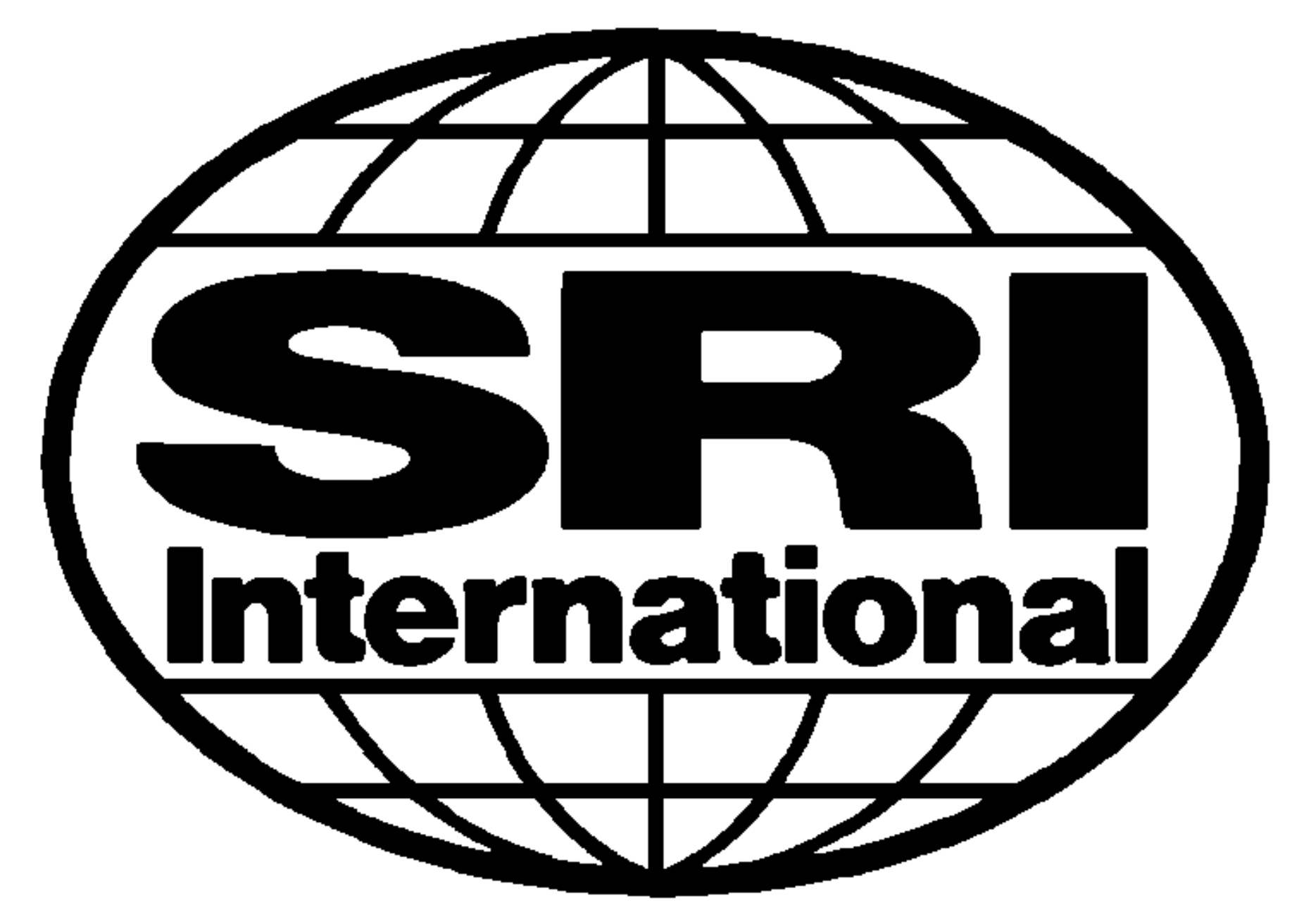}}

\newtheorem{theorem}{Theorem}

\newcommand{\defdef}{\mbox{$\stackrel{\rm def}{=}$}}
\def\defn{\mathrel{\defdef}}

\newcommand{\aless}{\mathrel{\mbox{\lower.9ex\hbox{$\stackrel{\textstyle <}{\sim}$}}}}
\newcommand{\amore}{\mathrel{\mbox{\lower.9ex\hbox{$\stackrel{\textstyle >}{\sim}$}}}}

\newcommand{\g}{\mathfrak{g}}
\newcommand{\wig}{\(\sim\)}

\newenvironment{sessionbiglab}[1]{\begin{flushleft}
 \setlength{\hsbw}{\linewidth}
 \addtolength{\hsbw}{-\arrayrulewidth}
 \addtolength{\hsbw}{-\tabcolsep}
 \begin{tabular}{@{}|c@{}|@{}}\hline 
 \begin{minipage}[b]{\hsbw}
 \vspace*{-.5pt}
 \begin{flushright}
 \rule{0.01in}{.15in}\rule{1.45in}{0.01in}\hspace{-1.5in}
 \raisebox{0.04in}{\makebox[1.45in][c]{\footnotesize #1}}
 \end{flushright}
 \vspace*{-.47in}
 \begingroup\small\vspace*{2.0ex}\begin{alltt}}{\end{alltt}\endgroup\end{minipage}\\ \hline 
 \end{tabular}
 \end{flushleft}}
\title{PVS Embeddings of Propositional\\[0.7ex]and Quantified Modal Logic}
\author{John Rushby\\
\emph{Computer Science Laboratory}\\
\emph{SRI International, Menlo Park CA USA}}
\begin{document}
\maketitle

\begin{abstract}

Modal logics allow reasoning about various \emph{modes} of truth: for
example, what it means for something to be \emph{possibly} true, or to
\emph{know} that something is true as opposed to merely
\emph{believing} it.  This report describes embeddings of
propositional and quantified modal logic in the PVS verification
system.  The resources of PVS allow this to be done in an attractive
way that supports much of the standard syntax of modal logic, while
providing effective automation.

The report introduces and formally specifies and verifies several
standard topics in modal logic such as relationships between the
standard modal axioms and properties of the accessibility relation,
and attributes of the Barcan Formula and its converse in both constant
and varying domains.

\end{abstract}

\tableofcontents

\listoffigures
\cleardoublepage

\section{Introduction}

The motivation for modal logics is to reason about various
\emph{modes} of truth: for example, what it means for something to be
\emph{possibly} true, or to \emph{know} that something is true as
opposed to merely \emph{believing} it.  The modal \emph{qualifier}
$\Box$ and its dual $\Diamond$ (defined as $\neg \Box \neg$) are used
to indicate expressions that should be interpreted modally.  All modal
logics share the same basic structure but they employ different sets
of axioms and make other adjustments according to the mode attributed
to the qualifiers.  For example, in an \emph{Epistemic} modal logic,
where $\Box$ is interpreted as knowledge, we will expect the formula
$\Box P \supset P$ to hold: if I \emph{know} that something is true,
then it should be true (a traditional definition of knowledge is
justified \emph{true} belief).  But we would not expect this formula to
hold in a \emph{Doxastic} logic, where $\Box$ is interpreted as
belief.  Instead, we might expect $\Box P \supset \Diamond P$ to hold:
if I \emph{believe} P is true, then I cannot also believe it to be
false (reading $\Diamond P$ as $\neg \Box \neg P$).  There is a
collection of common formulas such as these that have standard names
(the two above are called T and D, respectively) and that are used in
various combinations to axiomatize different modal logics.

Standard modal logics are obtained by enriching either propositional
logic or quantified (i.e,, first- or higher-order) logic to yield
propositional and quantified modal logics, respectively.  The
semantics of modal logics are derived from those of classical logics
by interpreting all expressions relative to a set of \emph{possible
worlds}.  Thus, whereas a constant $x$ has some fixed interpretation
in a classical logic, in a modal logic its interpretation depends on
the world $w$ and this can be encoded or \emph{embedded} in classical
logic by \emph{lifting} $x$ to a function on worlds: $x(w)$ then
denotes the interpretation of $x$ in world $w$.  The qualifier $\Box$
is interpreted as truth in \emph{all} possible worlds, so $\Box P$ is
$\forall w: P(w)$, while $\Diamond P$ applies to \emph{some} possible
worlds: $\exists w: P(w)$.  There are complexities in the details
because some worlds may not be \emph{accessible} from others, and it
is this that distinguishes the different modes from each other (and is
closely related to the choice of axioms such as T and D).

In the next section, I will describe more of propositional modal logic
and show how this can be embedded within PVS \cite{PVS}.  The
embedding uses the resources of PVS in a way that allows modal
formulas to be written in their standard syntax and used in
combination with the other features of PVS (for example, its rich type
system and comprehensive language for expressions and definitions).
Proofs of modal formulas use the standard capabilities of the PVS
prover and can be highly automated.

In the section following that, I describe quantified modal logic and
its embedding in PVS.   This embedding is only a slight extension of
that for propositional modal logic, but the combination of modal
logic and quantification has some intrinsic complexities that I
describe and illustrate.   In the final section, I discuss some topics
in the pragmatics of constructing specifications in quantified modal
logic, particularly the combination of modal and classical quantifiers.

\section{Propositional Modal Logic in PVS}

As mentioned in the introduction, the basic idea of the possible worlds
interpretation for propositional modal logic is that expressions of
propositional logic are \emph{lifted} and interpreted relative to
a possible world.  Thus, a generic expression $P$ in propositional
logic must be transformed into its lifted form $l(P)$, which is a
function that can be applied to a world $w$ to deliver its value
$l(P)(w)$ in that world.  The lifting transformation is defined
recursively over the constructs of propositional logic: that is, for a
propositional connective such as $\wedge$, we define how
$l(P\,\wedge\,Q)(w)$ is related to $l(P)(w)$ and $l(Q)(w)$, and
similarly for other constructs.  This is done below.

\begin{itemize}
\item Constants $a$ or variables $x$ (these are equivalent in
propositional logic), are lifted by a \emph{valuation} function $V$ where
$V(a)(w)$ and $V(x)(w)$ provide their values in world $w$.

\item Negation is lifted by negating the lifted term: $l(\neg\! P)(w)$ is
$\neg l(P)(w)$.

\item Conjunction is lifted by conjoining its lifted conjuncts: $l(P
\,\wedge\, Q)(w)$ is $l(P)(w) \,\wedge\, l(Q)(w)$.  The other binary
connectives are lifted in the same way.
\end{itemize}

We then specify the modal qualifiers as follows.
\begin{itemize}
\item $l(\Box P)$ is $\forall w: l(P)(w)$, where $w$ is a fresh variable.

\item $l(\Diamond P)$ is $\exists w: l(P)(w)$, where $w$ is a fresh
variable.\\ (An alternative, but equivalent, interpretation is that
$\Diamond P$ is $\neg \Box \neg P$)
\end{itemize}
Finally, a modal sentence $P$ is valid if it is true in all possible
worlds; that is, $\forall w: l(P)(w)$.

To illustrate these notions, I will use the following simple modal
argument.\footnote{This is actually a flawed version of Hartshorne's
rendition of Anselm's \emph{Proslogion} III argument for the existence
of God \cite[Section 4.1]{Eder&Ramharter15}.  The flaw is use of a
variable $P$ in H2 where the actual argument uses $g$.  Both forms of
the premise are questionable but the one used here is worse; I employ
it to illustrate aspects of propositional modal logic and its
embedding.  Those interested in the actual argument are referred to
\cite{Rushby21:ijpr}.}  This uses an \emph{Alethic} modal logic
where the qualifiers are interpreted as necessity ($\Box$) and
possibility ($\Diamond$).  Interpretation of the modal qualifiers is
related to an \emph{accessibility relation} on worlds and
corresponding axioms.  For simplicity of exposition, I delay these
topics to Section \ref{fullprop}, but the elementary treatment (i.e.,
without an accessibility relation) used in the subsections
before then is sound for Alethic logics (it is equivalent to the logic
known as S5).

\begin{description}\label{hartarg}
\item[Notation:] $g$ is a constant, $P$ is a metavariable.

\item[Premise H1:] $\Diamond g$ (i.e., $g$ is possible)

\item[Premise H2:] $P \supset \Box P$ (i.e., that which is true is
necessarily true).

\item[Conclusion HC:] $g$ (i.e., $g$ is true in the classical sense).
\end{description}

Each of these modal sentences is interpreted as the validity of its
lifted form, so premise H1 becomes $\forall w: l(\Diamond g)$, which
becomes $\forall w: \exists v: l(g)(v)$, which becomes $\forall w:
\exists v: V(g)(v)$, the translation shown below.  The outermost
quantifier is superfluous in this case.
\begin{description}
\item[H1:] $\forall w: \exists v: V(g)(v)$

\item[H2:] $\forall w: P(w) \supset (\forall v: P(v))$

\item[HC:] $\forall w: V(g)(w)$
\end{description}

H2 is similarly translated by recursively traversing its parse tree.
We start with modal validity of $P \supset \Box P$, which becomes
$\forall w: l(P \supset \Box P)(w)$; processing the $\supset$
connective, this becomes $ \forall w: l(P)(w) \supset l(\Box P)(w)$,
and then interpretation of the $\Box$ qualifier yields $\forall w:
(l(P)(w) \supset \forall v: l(P)(v))$; lifted metavariables are
themselves, so we end with the translation shown above.
HC likewise becomes $\forall w: l(g)(w)$, which becomes $\forall w:
V(g)(w)$, as shown above.

\begin{figure}[ht]
\begin{session}
eg_direct: THEORY
BEGIN
  worlds: TYPE+
  pmlformulas: TYPE = [worlds -> bool]
  pvars: TYPE+

  v, w: VAR worlds
  x: VAR pvars

  val(x)(w): bool
  
  g: pvars
  P: VAR pmlformulas

  H1: AXIOM EXISTS w: val(g)(w)

  H2: AXIOM P(w) IMPLIES FORALL v: P(v)

  HC: THEOREM val(g)(w)

END eg_direct
\end{session}
\caption{\label{shallowhart-pvs}Direct Shallow Embedding in PVS of
Propositional Example}
\end{figure}

Figure \ref{shallowhart-pvs} presents a direct transliteration into
PVS of the possible worlds interpretation of the example argument
that we constructed above.  The type \texttt{pmlformulas} (for
propositional modal logic formulas) is used for lifted formulas; its
variables (such as \texttt{P}) function as metavariables of the
embedded logic.  The type \texttt{pvars} is used for propositional
variables (and constants), with \texttt{val} as their valuation
function (cf. $V$ in the mathematical rendition above); \texttt{g} is
one of these propositional constants.  We then state the lifted
renditions of the premises and conclusion of the argument,
exploiting the fact that PVS automatically applies universal closure
to its formulas.

PVS proves the theorem HC automatically given the two premises.
\begin{sessionlab}{PVS proof}
(grind-with-lemmas :polarity? t :lemmas ("H1" "H2"))
\end{sessionlab}

An alternative way to state the theorem is to use an arbitrary world
``\texttt{here}.''

\begin{sessionlab}{PVS fragment}
  here: worlds

  HC_alt: THEOREM val(g)(here)
\end{sessionlab}
The same PVS proof as before will prove this alternative statement of
the theorem.

\subsection{Elementary Shallow Embedding of Propositional Modal Logic in PVS}
\label{elem-pml}

The kind of transformation from one logic or language to another seen
here is referred to as a \emph{shallow embedding}
\cite{Boulton92:embedding}.  The characteristic of a shallow embedding
is that it is a syntactic transformation: the modal presentation of
the example argument on page \ref{hartarg} is translated into the PVS
specification shown in Figure \ref{shallowhart-pvs}.  It would seem,
therefore, that automating the transformation will require a
syntax-to-syntax translator.  Fortunately, the capabilities of PVS
allow us to accomplish the translation quite effectively within PVS
itself.  This is feasible because the source language, propositional
modal logic, is a \emph{logic} and has much of its syntax in common
with PVS; the techniques we are about to see would be less effective if
the source were, say, a programming language.

The idea is to define the operators and connectives of the embedding
of propositional modal logic directly in their lifted form.  Thus,
whereas we earlier defined the lifted form of conjunction $l(P \,\wedge\,
Q)(w)$ to be $l(P)(w) \,\wedge\, l(Q)(w)$, here we will define a new modal
conjunction operator $\&_{m}$ by $(P \&_m Q)(w) = P(w) \wedge Q(w)$.  PVS
allows function symbols and names to be \emph{overloaded} (that is,
the same symbol or name can be used for several different functions)
and types are used to resolve the correct instance.  Thus, in PVS we
do not need a separate function name $\&_{m}$, we simply overload the
existing $\&$ (whose built-in definition is a synonym for Boolean
\texttt{AND}).\footnote{Likewise \wig\ is already
defined as a synonym for \texttt{NOT} and \texttt{=>} as a synonym for
\texttt{IMPLIES}; for ease of human parsing, we use symbols for the
lifted operators and ASCII for the Boolean ones.}

This is done in the theory \texttt{shallow\_pml} (for shallow
propositional modal logic) shown in Figure \ref{shallowpml1}.  The
first few blocks of declarations are the same as in
\texttt{direct\_hart}, then we define the lifted connectives
\wig, \texttt{\&}, and \texttt{=>} in the manner
described above.  Obviously, other lifted propositional connectives
can be added here in a similar way, or they can be defined in terms of
those already defined.  Notice that we are using symbols (e.g.,
\texttt{\&}) as function names here; the PVS Language reference
specifies the symbols that may be defined in this way \cite[Figure
2.4]{PVS:language}.

\begin{figure}[ht]
\begin{session}
elem_shallow_pml: THEORY
BEGIN

  worlds: TYPE+
  pmlformulas: TYPE = [worlds -> bool]
  pvars: TYPE+

  v, w: VAR worlds
  x, y: VAR pvars
  P, Q: VAR pmlformulas

  val(x)(w): bool

  \(\sim\)(P)(w): bool = NOT P(w) ;
  &(P, Q)(w): bool = P(w) AND Q(w) ;
  =>(P, Q)(w): bool = P(w) IMPLIES Q(w) ;
  □(P)(w): bool = FORALL v: P(v) ;

  <>(P): pmlformulas = \(\sim\) □ \(\sim\) P ;

  |=(w, P): bool = P(w)
  valid(P): bool = FORALL w: w |= P

END elem_shallow_pml
\end{session}
\caption{\label{shallowpml1}Elementary Shallow Embedding of
Propositional Modal Logic in PVS}
\end{figure}

Next, we define the $\Box$ qualifier of modal logic.  The PVS Language
reference suggests we could use \texttt{[]} here (as we use
\texttt{<>} below), but recent versions of PVS preempt this syntax for
declaration-level parameterization.  However, these recent versions of
PVS also allow use of Unicode, so we simply employ the appropriate
Unicode character (\texttt{□}, hexadecimal code \texttt{25A1}) as the
name of our function.  Next, we define \texttt{<>} (we could have used
\texttt{◇}, Unicode \texttt{25C7} instead) as the dual modal
qualifier.  We could have done this in a similar way to \texttt{□}
(but using an existential quantifier), but for variety we will instead
define it in terms of \texttt{□}.  Notice that no parentheses are used
here (i.e., we do not need
\texttt{\wig(□(\wig(P)))}).  This is because
\texttt{\wig}, \texttt{□}, and \texttt{<>} are known to be
\emph{unary} operators \cite[Figure 7.1]{PVS:language}.\footnote{My
actual recommendation is to use the definition \texttt{<>(P)(w): bool
= EXISTS v: P(v)} because it is much easier to interpret the
existential quantifier than the doubly negated universal when guiding
interactive PVS proofs.}  For symmetry with later constructions, we
define \texttt{w |= P} to be truth of \texttt{P} in world \texttt{w}.
Notice that \texttt{|=}, together with \texttt{\&} and \texttt{=>},
are known to PVS as binary infix operators \cite[Figure
7.1]{PVS:language}, although they must appear in prefix form when
being defined.  Finally, we define modal validity in the expected way.

Now that we have a shallow embedding of propositional modal logic in
PVS, we can simply import it into a new theory where we specify
the example argument in a fairly direct way.  This is shown in Figure
\ref{shallowhart-pvs2}; the theorem is proved automatically in the
same way as before.  Notice that the modalities \texttt{□} and
\texttt{<>} are used here in a natural way because, as noted before,
PVS treats these symbols as unary operators.  

\begin{figure}[ht]
\begin{session}
eg_elem_shallow1: THEORY
BEGIN IMPORTING elem_shallow_pml

  g: pvars
  P: var pmlformulas

  H1: AXIOM valid(<> val(g))
 
  H2: AXIOM valid(P => □ P)

  HC: THEOREM valid(val(g))

END eg_elem_shallow1
\end{session}
\caption{\label{shallowhart-pvs2}Partially Automated Shallow PVS Embedding of
Propositional Example}
\end{figure}

All the formulas in Figure \ref{shallowhart-pvs2} explicitly employ
the function \texttt{valid} to reduce validity of modal sentences to
the classical validity employed in PVS\@.  It would be nice to
automate this and PVS provides a way to do it: we specify that
\texttt{valid} is a \texttt{CONVERSION}.  When a PVS expression fails
to typecheck, PVS searches for a conversion function that will make it
type-correct and applies it automatically (PVS provides commands that
prettyprint the specification with conversions applied, and these are
also shown expanded in proofs).  Use of the conversion allows
\texttt{H2} to be written in the standard modal syntax.  If we
additionally specify \texttt{val} as a conversion then \texttt{H1}
also can be written in the standard syntax.  However, the conclusion
\texttt{HC} still requires verbose syntax: we would like to write just
\texttt{g}.  The reason we cannot do this is that it requires
application of two conversions before \texttt{g} becomes type correct.
We can define a function \texttt{validval} that applies both
\texttt{valid} and \texttt{val} and declare that to be a conversion.
Use of all three conversions allows us to simplify the PVS
specification of the example argument to the form shown in Figure
\ref{shallowhart-pvs3}.

\begin{figure}[ht]
\begin{session}
eg_elem_shallow2: THEORY
BEGIN IMPORTING elem_shallow_pml

  g: pvars
  P: var pmlformulas

  validval(x: pvars): bool = valid(val(x))
  CONVERSION valid, val, validval

  H1: AXIOM <> g

  H2: AXIOM P => □ P

  HC: THEOREM g

END eg_elem_shallow2
\end{session}
\caption{\label{shallowhart-pvs3}More Automated Shallow PVS Embedding of
Propositional Example}
\end{figure}

It will generally be more convenient to specify these conversions in
the theory \texttt{elem\_shallow\_pml} as they will then automatically
be available wherever this is imported.  The theorem \texttt{HC} is
proved automatically in the same way as before---for, semantically, it
is the same as the previous versions; the automation simply allows
better syntax.

Notice that if PVS did not allow overloading of built-in and infix
symbols, then the middle block of definitions in Figure
\ref{shallowpml1} would have to be written using standard functional
notation as follows.
\begin{sessionlab}{PVS fragment}
  mneg(P)(w): bool = NOT P(w) ;
  mand(P, Q)(w): bool = P(w) AND Q(w) ;
  mimp(P, Q)(w): bool = P(w) IMPLIES Q(w) ;
  mbox(P)(w): bool = FORALL v: P(v) ;
  mdia(P): pmlformulas = mneg(mbox(mneg(P)))
\end{sessionlab}
The premises of the example argument would then appear like this.
\begin{sessionlab}{PVS fragment}
  H1: AXIOM valid(mdia(val(g)))
 
  H2: AXIOM valid(mimp(P, mbox(P)))
\end{sessionlab}
The improvement in Figure \ref{shallowhart-pvs3} is obvious.

Now that we know of \emph{shallow} embeddings, it will come as no
surprise that there are \emph{deep} embeddings also.  I introduce
these in the next subsection.

\subsection{Elementary Deep Embedding of Propositional Modal Logic in PVS}

Whereas shallow embeddings work on the surface syntax of the source
language, deep embeddings are defined on its abstract syntax: we
define an abstract datatype that models this syntax, then define
functions that operate on it by recursion and case analysis
\cite{Boulton92:embedding}.  The abstract syntax for propositional
modal logic is specified by the PVS datatype in Figure \ref{pml-adt}.
\begin{figure}[ht]
\begin{session}
modalformula[pvars: TYPE+]: DATATYPE
BEGIN
  pvar(arg: pvars): var?
  \(\sim\)(arg: modalformula): not?
  &(arg1: modalformula, arg2: modalformula): and?
  =>(arg1: modalformula, arg2: modalformula): imp?
  □(arg: modalformula): box?
END modalformula
\end{session}
\caption{\label{pml-adt}PVS Datatype for Abstract Syntax of
Propositional Modal Logic}
\end{figure}

This abstract datatype defines the \texttt{modalformula} datatype,
parameterized by the nonempty type \texttt{pvars} that specifies its
propositional variables.  We then import this into the theory
\texttt{elem\_deep\_pml} shown in in Figure \ref{deep-pml} where we define
validity of \texttt{modalformula} \texttt{P} in world \texttt{w},
\texttt{w |= P}, by recursion on the structure of \texttt{P}\@.  PVS
ensures termination of recursive definitions by generating proof
obligations and the \texttt{MEASURE} keyword instructs PVS to use the
subterm ordering relation \texttt{<<} in this proof obligation.  The
rest of the construction is the same as in \texttt{shallow\_pml}.

We can then employ \texttt{elem\_deep\_pml} in the representation of
the example argument shown in Figure \ref{deep-hart}.  Notice that
syntactically this identical to the representation using a shallow
embedding that we saw in Figure \ref{shallowhart-pvs3}.  

\begin{figure}[ht!]
\begin{session}
eg_elem_deep: THEORY
BEGIN IMPORTING elem_deep_pml

  g: pvars
  P: VAR modalformula

  H1: AXIOM <> g

  H2: AXIOM P => □ P

  HC: THEOREM g

END eg_elem_deep
\end{session}
\caption{\label{deep-hart}Representation of the Propositional Example
    Using Deep Embedding in PVS}
\end{figure}

\begin{figure}[ht]
\begin{session}
elem_deep_pml: THEORY
BEGIN
  
  worlds, pvars: TYPE+
  IMPORTING modalformula[pvars]

  w, v: VAR worlds
  x, y: VAR pvars
  P, Q: VAR modalformula

  val(x)(w): bool

  |=(w, P): RECURSIVE bool =
     CASES P OF
      pvar(x): val(x)(w),
      \(\sim\)(R): NOT (w |= R),
      &(R, S): (w |= R) AND (w |= S),
      =>(R, S): (w |= R) IMPLIES (w |= S),
      □(R): FORALL v: (v |= R)
    ENDCASES
  MEASURE P by <<

  <>(P): modalformula = \(\sim\) □ \(\sim\) P ;

  valid(P): bool = FORALL w: w |= P
  validval(x: pvars): bool = valid(pvar(x))
  CONVERSION valid, pvar, validval

END elem_deep_pml
\end{session}
\caption{\label{deep-pml}Elementary Deep Embedding of Propositional
Modal Logic in PVS}
\end{figure}

The proof of the theorem is shown below.  It installs the two
premises, instantiates one of them, then applies the standard
top-level proof strategy of PVS.  This is not quite as automated as
the proof using a shallow embedding but this is due to the rather weak
quantifier instantiation heuristics in PVS rather than intrinsic
difficulty.
\begin{sessionlab}{PVS proof}
(lemma "H1") (lemma "H2") (inst?) (grind :polarity? t)
\end{sessionlab}

For the purposes described here, there is little to choose between
shallow and deep embeddings.  For the remainder of this tutorial I
will mostly use the shallow embedding, because it is somewhat simpler.
A deep embedding would be preferable if we wanted to establish
metalogical properties (i.e., properties \emph{about} modal logic as
opposed to properties stated \emph{in} modal logic).

All
our embeddings have so far implicitly assumed that all possible worlds
are accessible from each other.  This is an
oversimplification\footnote{Actually, it does correspond to a
legitimate modal logic---a variant on S5---that is described later.}
that we correct in the following section.

\subsection{Shallow Embedding of Full Propositional Modal Logic in PVS}

\label{fullprop}

The modal qualifiers allow us to speak about truth in other possible
worlds: when we say $\Diamond P$ we are saying that there are possible
worlds where $P$ is true.  But suppose there is a structure on
possible worlds and we cannot reach every possible world from every
other one; then it might be that $P$ is true only in worlds we cannot
reach, so $\Diamond P$ would be false in this configuration of worlds.
A full formalization of propositional modal logic uses an
\emph{accessibility relation} on possible worlds and adjusts the
semantics of the modal qualifiers so they apply only to accessible
worlds.  In the shallow PVS embedding of Figure \ref{shallowpml1}, we
do this as follows, where \texttt{access} is a new declaration that
introduces the accessibility relation and the definition of
\texttt{$\Box$} is modified to reference this.  There is no need to
modify the definition of \texttt{<>} if this is given in terms of
\texttt{$\Box$}, but if it is defined directly, it takes the form
shown below.   We can make exactly analogous adjustments to the
deep embedding.

\begin{sessionlab}{PVS fragment}
  access(w, v): bool

 \, □\,(P)(w): bool = FORALL v: access(w, v) IMPLIES P(v) ;
  <>(P)(w): bool = EXISTS v: access(w, v) AND P(v) ;
\end{sessionlab}

To complete our full shallow and deep PVS embeddings of propositional
modal logic, we need to parameterize the theories.  Currently the type
\texttt{worlds}, its accessibility relation \texttt{access}, and the
type \texttt{pvars} and its valuation function \texttt{val} are
defined within the embedding theories.  We need instead to specify some
or all of these these as parameters, so that they can be defined by
the theories that use the embedding.  It is a matter of choice which
of these types and constants are defined as parameters, but there are
some dependencies.  Certainly \texttt{pvars}, the type of the
propositional variables, should almost certainly be a parameter.  We
will see a circumstance later where \texttt{val}, the valuation
function needs to be a parameter; its type is \texttt{[pvars ->
[worlds -> bool]]} so this forces both \texttt{pvars} and
\texttt{worlds} to be parameters, in which case we may as well
complete the parameterization by adding \texttt{access}.

\begin{figure}[ht]
\begin{session}
full_shallow_pml [worlds: TYPE+, access: pred[[worlds,worlds]],
               pvars: TYPE+, val: [pvars -> [worlds -> bool]]]: THEORY
BEGIN

  pmlformulas: TYPE = [worlds -> bool]
  v, w: VAR worlds
  x, y: VAR pvars
  P, Q: VAR pmlformulas

  \wig(P)(w): bool = NOT P(w) ;
  &(P, Q)(w): bool = P(w) AND Q(w) ;
  =>(P, Q)(w): bool = P(w) IMPLIES Q(w) ;
 
\,\,  □\,(P)(w): bool = FORALL v: access(w, v) IMPLIES P(v) ;
  <>(P)(w): bool = EXISTS v: access(w, v) AND P(v) ;

  |=(w, P): bool = P(w)
  valid(P): bool = FORALL w: w |= P

  validval(x: pvars): bool = valid(val(x))
  CONVERSION valid, val, validval

END full_shallow_pml
\end{session}
\caption{\label{full-shallow}Full Shallow Embedding of Propositional Modal
    Logic in PVS}
\end{figure}

This is shown in Figure \ref{full-shallow}, which is a fully
parameterized version of the shallow embedding for full propositional
modal logic.  We name this modified PVS specification
\texttt{full\_shallow\_pml} and use it in the version of the example
argument shown in Figure \ref{fullshal-hart}.  The statement of the
theorem is adjusted to record the fact that it requires the
accessibility relation to be symmetric.  The
\texttt{full\_shallow\_pml} theory specifies \texttt{valid},
\texttt{val}, and \texttt{validval} as conversions.

\begin{figure}[ht]
\begin{session}
eg_full_shallow: THEORY
BEGIN

  worlds, pvars: TYPE+
  access: pred[[worlds, worlds]]
  val(x:pvars)(w:worlds): bool  

  IMPORTING full_shallow_pml[worlds, access, pvars, val]

  g: pvars
  P: var pmlformulas

  H1: AXIOM <> g

  H2: AXIOM P => □ P

  HC: THEOREM symmetric?(access) => g

END eg_full_shallow
\end{session}
\caption{\label{fullshal-hart}The Propositional Example Using an
Embedding of Full Propositional Modal Logic}
\end{figure}

As in the previous version, the proof requires some manual steps to
guide the quantifier instantiation.

\begin{sessionlab}{PVS proof}
(grind-with-lemmas :lemmas ("H1" "H2"))
(inst -1 "val(g)")
(inst? -2)
(grind :polarity? t)
\end{sessionlab}

There is a correspondence between properties of the accessibility
relation and certain modal formulas: for example, there is a modal
formula that exactly corresponds to the accessibility relation being
symmetric.  We examine this topic in the next section.

\subsection{Standard Axioms and Various Modal Logics}
\label{stdax}

As we noted in the introduction, different modal logics apply
different interpretations to the modal qualifiers and employ different
sets of axioms to characterize their properties.  There is a
collection of standard axioms with single-character names that are
used in various combinations to axiomatize different modal logics.
The most widely used axioms are the following.

\begin{description}
\item[K:] $\Box(P \supset Q) \supset (\Box P \supset \Box Q)$,
  
\item[T:] $\Box P \supset P$, 
\item[4:] $\Box P \supset \Box \Box P$,
\item[B:] $P \supset \Box\Diamond P$,
\item[D:] $\Box P \supset \Diamond P$,
\item[5:] $\Diamond P \supset \Box\Diamond P$.
\end{description}
In addition to these axioms, there is a principle of
\emph{necessitation}, generally named N.
\begin{description}
\item[N:] if $P$ is valid, so is $\Box P$.
\end{description}
Necessitation is actually a metatheorem, true in all modal logics; it
says that if $P$ is just plain true (i.e., without reference to a
possible world), then it is true in all possible worlds.

The different modal logics incorporate different sets of axioms, which
are indicated by juxtaposing their names; thus a standard Deontic
(obligation) logic uses \textbf{KD}, Doxastic (belief) logic uses
\textbf{KD45}, while Epistemic (knowledge), and Alethic (necessity)
logics use \textbf{KT45}, which is also known as
\textbf{S5}.\footnote{S5 differs from the modal logic introduced in
Section \ref{elem-pml} (where the accessibility relation was absent) in
that its worlds can be structured as several isolated cliques, whereas
in the logic of the earlier section all worlds implicitly belong to a
single clique.  However, the two logics have the same valid sentences
and can be considered equivalent.}  Elementary logics of time use
\textbf{KT4}, which is also known as \textbf{S4}, but the temporal
logics in computer science such as LTL and CTL have rather more
structure (they add a \emph{next} operator).  Notice that \textbf{K}
is always present: it is actually a theorem, true in all modal logics,
rather than an axiom.

It is a remarkable fact that each of the axioms above corresponds to a
fairly natural property of the accessibility relation (apart from N and K,
which are always true).
\begin{description}
\item[T:] the accessibility relation is reflexive
\item[4:] the accessibility relation is transitive
\item[B:] the accessibility relation is symmetric
\item[D:] the accessibility relation is serial
\item[5:] the accessibility relation is Euclidean
\end{description}

If we denote the accessibility relation by $R$, the serial property is
$\forall w: \exists v: R(w, v)$ while Euclidean is $\forall u, v, w:
R(u, v) \wedge R(u, w) \supset R(v, w)$.  A relation that is both
symmetric and Euclidean is also transitive, and a relation that is
both reflexive and Euclidean is also symmetric and hence transitive
and, therefore, an equivalence relation.  Thus, the accessibility relations of
modal logics that include axioms T and 5 are equivalence relations.

Another concept that uses modal constructs is \emph{strict}
implication, usually written $\strictif$, in contrast to the
\emph{material} implication $\supset$ of classical logic.  We say that
$P$ strictly implies $Q$ if it is not possible for $P$ to be true and
$Q$ false: that is, in an Alethic modal logic
$$ P \strictif Q  \defn \neg \Diamond (P \wedge \neg Q).$$
It is a theorem of Alethic logic that strict implication is the same
as necessary material implication:
$$ P \strictif Q = \Box (P \supset Q).\footnote{This equality is a
theorem of all modal logics (i.e., it requires no axioms) but it carries
the intended interpretation only in Alethic logics.}$$

We will now add these notions to the PVS embeddings of propositional
modal logic and prove the various theorems.  We use the shallow embedding
for illustration, but the deep embedding is handled similarly.

First, we define strict implication (as the infix symbol \texttt{|>})
and the standard modal axioms.  These are shown in Figure
\ref{modal-axioms}, preceded by a theory that defines serial and
Euclidean relations and their properties, as these are lacking from
the \texttt{relations} theory built in to the PVS Prelude.  All the
lemmas in these theories are proved automatically, apart from the very
first, \texttt{sym\_Euc}, which requires a manual quantifier
instantiation step.

\begin{figure}[htp]
\begin{session}
more_relations [T: TYPE]: THEORY
BEGIN

  IMPORTING relations[T]
  R: VAR PRED[[T, T]]
  x, y, z: VAR T

  serial?(R):    bool = FORALL x: EXISTS y: R(x, y)
  Euclidean?(R): bool = FORALL x, y, z: R(x, y) & R(x, z) => R(y, z)

  sym_Euc:  LEMMA symmetric?(R) AND Euclidean?(R) IMPLIES transitive?(R)
  ref_Euc1: LEMMA reflexive?(R) AND Euclidean?(R) IMPLIES symmetric?(R)
  ref_Euc2: LEMMA reflexive?(R) AND Euclidean?(R) IMPLIES equivalence?(R)

  equiv_is_Euclidean: JUDGEMENT (equivalence?) SUBTYPE_OF (Euclidean?)

END more_relations

modal_axioms: THEORY
BEGIN

  worlds, pvars: TYPE+
  val: [pvars -> [worlds -> bool]]
  access: pred[[worlds,worlds]]
  IMPORTING full_deep_pml [worlds, access, pvars, val]
  IMPORTING more_relations[worlds]

  P, Q: VAR pmlformulas

  |>(P, Q): pmlformulas = \(\sim\)<>(P & \(\sim\)Q)

  strict_material: LEMMA P |> Q = □(P => Q)

  K: LEMMA □(P => Q) IMPLIES (□P => □Q)
  N: LEMMA valid(P) IMPLIES valid(□P)

  T: AXIOM □ P => P
  four: AXIOM □ P => □ □ P
  B: AXIOM P => □ <> P
  D: AXIOM □ P => <> P
  five: AXIOM <> P => □ <> P

END modal_axioms
\end{session}
\caption{\label{modal-axioms}Strict Implication and the Standard Modal
Axioms in PVS}
\end{figure}

Some remarks on the modal axioms seem appropriate at this point.
Those that are used in standard theories (i.e., TD45) seem appropriate
to their intended interpretation; B is more controversial, however.
It looks innocuous, but it is equivalent to $\Diamond \Box P \supset
P$, which seems less so.  Premises similar to H2 are generally
justified by virtue of their similarity to N, the principle of
necessitation, which is true in all modal logics.  That similarity is
superficial, however.  If we invite PVS to expand N in its shallow
embedding from the theory \texttt{modal\_axioms} of Figure
\ref{modal-axioms}, it delivers the following
\begin{sessionlab}{N in PVS}
(FORALL u: P!1(u))
       IMPLIES (FORALL w: FORALL v: access(w, v) IMPLIES P!1(v))
\end{sessionlab}
whereas doing the same for H2 in \texttt{eg\_full\_shallow} of Figure
\ref{fullshal-hart} delivers the following.
\begin{sessionlab}{H2 in PVS}
(FORALL w: P!1(w)
       IMPLIES (FORALL v: access(w, v) IMPLIES P!1(v)))
\end{sessionlab}
The crucial difference in quantification (and parenthesization) is now
quite evident.  N says that if P is true in every world, then it must
be true in every world that is accessible from a given world.  This is
obviously true (and provable).  Whereas H2 says that if P is true in a
given world, it must also be true in every world that is accessible
from that world.  This is a strong claim and one that seems difficult
to justify in general.

This highlights another item worth noting, namely, that the deduction
theorem is not valid in modal logic.  That is to say, the following
(which would allow \texttt{H2} to be derived from \texttt{N}) is not
provable (its converse is, however).\label{nonded}
\begin{sessionlab}{PVS fragment}
  nondeduction: CLAIM (valid(P) IMPLIES valid(Q)) IMPLIES valid(P => Q)
\end{sessionlab}

Related to this is the observation that it is often necessary to be
careful about which parts of a sentence are to be interpreted modally,
and which are conventional propositional logic.  Specifically, a
correct statement of modal \emph{modus tollens} is expressed in PVS as
follows (i.e., two modal sentences connected propositionally),
\begin{sessionlab}{PVS fragment}
  tollens: LEMMA (P => Q) IMPLIES (□\wig{}Q => □\wig{}P)
\end{sessionlab}
whereas the following (i.e., a single modal sentence) is invalid.
\begin{sessionlab}{PVS fragment}
  tollens_bad: CLAIM (P => Q) => (□\wig{}Q => □\wig{}P)
\end{sessionlab}

Next, we prove that the standard modal axioms are each equivalent to a
property of the accessibility relation.  We do this in two stages.
First we prove that properties of the accessibility relation entail
the corresponding modal axiom.  This is done in the theory
\texttt{modal\_props} shown in Figure \ref{props-axioms}.
The formal specification here is a rather unsatisfactory.   What we
would like to say is something like the following
\begin{sessionlab}{Formerly not PVS}
  T_refl: LEMMA reflexive?(access) IMPLIES T
\end{sessionlab}
where we reference the name of one formula (here, \texttt{T}) within
another formula.  Unfortunately, PVS does not support
this,\footnote{Recent versions of PVS do support this, so
\texttt{T\_refl} is now valid PVS.}  so we have
to repeat the actual formula named by \texttt{T}.  All the lemmas are
proved automatically, except the last two, which require manual
quantifier instantiation.

\begin{figure}[ht]
\begin{session}
modal_props: THEORY
BEGIN

  IMPORTING modal_axioms
  IMPORTING more_relations[worlds]

  P: VAR pmlformulas

  T_refl: LEMMA reflexive?(access) IMPLIES □ P => P

  four_trans: LEMMA transitive?(access) IMPLIES □ P => □ □ P

  B_sym: LEMMA symmetric?(access) IMPLIES P => □ <> P

  D_serial: LEMMA serial?(access) IMPLIES □ P => <> P

  five_Euc: LEMMA Euclidean?(access) IMPLIES <> P => □ <> P

END modal_props
\end{session}
\caption{\label{props-axioms}Accessibility Relation Properties and the Standard Modal
Axioms in PVS}
\end{figure}

In the final stage, we prove that each of the modal axioms entails a
property of the accessibility relation (i.e., the reverse direction of
the previous stage).   By comparison with the previous theory, it
might seem that we would state these lemmas in a similar manner, as
shown below.
\begin{sessionlab}{PVS fragment}
  refl_T: LEMMA □ P => P IMPLIES reflexive?(access)
\end{sessionlab}

However there is a complication.  To introduce this, note that
the informal way to prove the result is via its contrapositive: we
suppose the accessibility relation is not reflexive, so there is some world
\texttt{w} such that \texttt{NOT access(w, w)}.  Now define the
valuation function \texttt{val} such that \texttt{val(p)(v) = NOT v=w}
for some \texttt{pvar} \texttt{p}.  Then \texttt{w |= pvar(p)} is
false but \texttt{w |= □ pvar(p)} is true (because the valuation of
\texttt{p} is true everywhere except at \texttt{w} and \texttt{w} is
not accessible from \texttt{w}).  Thus, we contradict the antecedent
and by \emph{reducto} conclude that the accessibility relation must be
reflexive.  This is a sound proof because \texttt{□ P => P} must be
true in all models.

\begin{figure}[ht]
\begin{session}
more_modal_props: THEORY
BEGIN

  worlds, pvars: TYPE+
  val: VAR [pvars -> [worlds -> bool]]
  access: pred[[worlds,worlds]]
  IMPORTING full_deep_pml
  IMPORTING more_relations[worlds]

  P: VAR modalformula[pvars]

  refl_T: LEMMA (FORALL P, val:
     full_deep_pml[worlds,access,pvars,val].valid(□P => P))
    IMPLIES reflexive?(access)

END more_modal_props
\end{session}
\caption[Accessibility Relation Properties and the Standard Modal Axioms\newline (other direction) in PVS]{\label{axioms-props}Accessibility Relation Properties and the
Standard Modal Axioms\newline\hspace*{0.7in}(other direction) in PVS}
\end{figure}

Now the issue in constructing this proof in PVS is that we need to be
able to define a suitable valuation function during the proof: thus
the function needs to be a variable.  In all our specifications so far
we have defined the parameters to the embedding as uninterpreted
constants.  Here, we need to keep the valuation function \texttt{val}
as a free variable---hence, the quantification in the specification of
the formula \texttt{refl\_T} shown in Figure \ref{axioms-props}.
Unlike earlier specifications, the conversion \texttt{valid} cannot be
applied automatically as PVS cannot tell what theory instance is
required, so we have to specify it in its qualified form, where we
state the full theory theory instance.  It is likewise difficult to
specify the type of \texttt{P}.  We cannot supply \texttt{val} as a
parameter to its theory instance as this is a variable, so we need to
reparameterize the embedding so that modal formulas are specified
separately from their interpretation.  Rather than do this for
\texttt{full\_shallow\_pml}, we use \texttt{full\_deep\_pml} as it is
already structured and parameterized in this way.

The proof of \texttt{refl\_T} is shown below.  The \texttt{(lemma
"pvars\_nonempty")} is used to access the name (\texttt{x!2}) of some
member of \texttt{pvars} (since this type is specified to be
nonempty).  The \texttt{inst} command then constructs the valuation
function described above, after which PVS can complete the proof
automatically.
\begin{sessionlab}{PVS proof}
(ground)
(expand "reflexive?")
(skosimp)
(lemma "pvars_nonempty")
(skosimp)
(inst - "pvar(x!2)"
  "LAMBDA (x:pvars): LAMBDA (w:worlds): NOT (x=x!2 AND w=x!1)")
(grind)
\end{sessionlab}

The theorems and proofs for the other modal axioms are constructed
similarly, but are sufficiently \sout{tedious} challenging that they
are left as exercises for the reader.  Another exercise is to rework
the version of the example argument in Figure \ref{fullshal-hart} so
that it cites modal Axiom B instead of symmetry of the accessibility
relation.

\subsection{\LaTeX-Printing}

PVS is able to typeset specifications in \LaTeX\ and thereby reproduce
standard mathematical notation.  We illustrate this on the following
example specification that uses strict implication.

\begin{sessionbiglab}{Example Specification}
  |>(P, Q): pmlformulas =  \wig\,<>(P & \wig\,Q)

  H1: AXIOM g |> □\,g

  H_triv: LEMMA symmetric?(access) IMPLIES (P |> □\,P) IMPLIES (<>\,P => P)

  H2: AXIOM <>\,g

  HC: THEOREM symmetric?(access) => g
\end{sessionbiglab}

\newpage
A \LaTeX-printed version of this example is shown below.

\def\munderscoretimestwofn#1#2{{#1 \times #2}}
\def\fmunderscoretimestwofn#1#2{{#1 \times #2}}
\def\sigmaunderscoretimestwofn#1#2{{#1 \times #2}}
\def\generatedunderscoresubsetunderscorealgebraonefn#1{{{\cal A}(#1)}}
\def\generatedunderscoresigmaunderscorealgebraonefn#1{{{\cal S}(#1)}}
\def\aeunderscoredecreasingotheronefn#1{{\pvsid{decreasing?}(#1)~\mbox{\it a.e.}}}
\def\aeunderscoreincreasingotheronefn#1{{\pvsid{increasing?}(#1)~\mbox{\it a.e.}}}
\def\aeunderscoreconvergenceothertwofn#1#2{{#1 \longrightarrow #2~\mbox{\it a.e.}}}
\def\aeunderscoreeqothertwofn#1#2{{#1 = #2~\mbox{\it a.e.}}}
\def\aeunderscoreleothertwofn#1#2{{#1 \leq #2~\mbox{\it a.e.}}}
\def\aeunderscoreposotheronefn#1{{#1> 0~\mbox{\it a.e.}}}
\def\aeunderscorenonnegotheronefn#1{{#1 \geq 0~\mbox{\it a.e.}}}
\def\aeunderscorezerootheronefn#1{{#1 = 0~\mbox{\it a.e.}}}
\def\xunderscorelttwofn#1#2{{#1 < #2}}
\def\xunderscoreletwofn#1#2{{#1 \leq #2}}
\def\xunderscoreeqtwofn#1#2{{#1 = #2}}
\def\xunderscoretimestwofn#1#2{{#1 \times #2}}
\def\xunderscoreaddtwofn#1#2{{#1 + #2}}
\def\xunderscorelimitonefn#1{{\pvsid{limit}(#1)}}
\def\xunderscoresumonefn#1{{\sum #1}}
\def\xunderscoresigmathreefn#1#2#3{{\sum_{#1}^{#2} #3}}
\def\xunderscoresuponefn#1{{\pvsid{sup}(#1)}}
\def\xunderscoreinfonefn#1{{\pvsid{inf}(#1)}}
\def\pointwiseunderscoreconvergesunderscoredowntoothertwofn#1#2{{#1 \searrow #2}}
\def\pointwiseunderscoreconvergesunderscoreuptoothertwofn#1#2{{#1 \nearrow #2}}
\def\pointwiseunderscoreconvergenceothertwofn#1#2{{#1 \longrightarrow #2}}
\def\convergesunderscoredowntoothertwofn#1#2{{#1 \searrow #2}}
\def\convergesunderscoreuptoothertwofn#1#2{{#1 \nearrow #2}}
\def\convergenceothertwofn#1#2{{#1 \longrightarrow #2}}
\def\convergencetwofn#1#2{{#1 \longrightarrow #2}}
\def\crossunderscoreproducttwofn#1#2{{#1 \times #2}}
\def\conjugateonefn#1{{\overline{#1}}}
\def\cunderscoredivtwofn#1#2{{#1/#2}}
\def\cunderscoremultwofn#1#2{{#1\times#2}}
\def\cunderscoresubtwofn#1#2{{#1-#2}}
\def\cunderscorenegonefn#1{{-#1}}
\def\cunderscoreaddtwofn#1#2{{#1+#2}}
\def\Imonefn#1{{\Im(#1)}}
\def\Reonefn#1{{\Re(#1)}}
\def\Etwofn#1#2{{\mathbb{E}(#1~|~#2)}}
\def\Eonefn#1{{\mathbb{E}(#1)}}
\def\Ptwofn#1#2{{\mathbb{P}(#1~|~#2)}}
\def\Ponefn#1{{\mathbb{P}(#1)}}
\def\xtwofn#1#2{{#1\times#2}}
\def\asttwofn#1#2{{#1\ast#2}}
\def\minusonefn#1{{{#1}^{-}}}
\def\plusonefn#1{{{#1}^{+}}}
\def\astonefn#1{{{#1}^{\ast}}}
\def\dottwofn#1#2{{#1\bullet#2}}
\def\integralthreefn#1#2#3{{\int_{#1}^{#2} #3}}
\def\integraltwofn#1#2{{\int_{#1} #2}}
\def\integralonefn#1{{\int#1}}
\def\normonefn#1{{\left||{#1}\right||}}
\def\phionefn#1{{\pvssubscript{\phi}{#1}}}
\def\infunderscoreclosedonefn#1{{\left(-\infty,~#1\right]}}
\def\closedunderscoreinfonefn#1{{\left[#1,~\infty\right)}}
\def\infunderscoreopenonefn#1{(-\infty,~#1)}
\def\openunderscoreinfonefn#1{(#1,~\infty)}
\def\closedtwofn#1#2{{\left[#1,~#2\right]}}
\def\opentwofn#1#2{(#1,~#2)}
\def\sigmathreefn#1#2#3{{\sum_{#1}^{#2} #3}}
\def\sigmatwofn#1#2{{\sum_{#1} {#2}}}
\def\ceilingonefn#1{{\lceil{#1}\rceil}}
\def\flooronefn#1{{\lfloor{#1}\rfloor}}
\def\absonefn#1{{\left|{#1}\right|}}
\def\roottwofn#1#2{{\sqrt[#2]{#1}}}
\def\sqrtonefn#1{{\sqrt{#1}}}
\def\sqonefn#1{{\pvssuperscript{#1}{2}}}
\def\expttwofn#1#2{{\pvssuperscript{#1}{#2}}}
\def\opcarettwofn#1#2{{\pvssuperscript{#1}{#2}}}
\def\indexedunderscoresetsotherIIntersectiononefn#1{{\bigcap #1}}
\def\indexedunderscoresetsotherIUniononefn#1{{\bigcup #1}}
\def\setsotherIntersectiononefn#1{{\bigcap #1}}
\def\setsotherUniononefn#1{{\bigcup #1}}
\def\setsotherremovetwofn#1#2{{(#2 \setminus \{#1\})}}
\def\setsotheraddtwofn#1#2{{(#2 \cup \{#1\})}}
\def\setsotherdifferencetwofn#1#2{{(#1 \setminus #2)}}
\def\setsothercomplementonefn#1{{\overline{#1}}}
\def\setsotherintersectiontwofn#1#2{{(#1 \cap #2)}}
\def\setsotheruniontwofn#1#2{{(#1 \cup #2)}}
\def\setsotherstrictunderscoresubsetothertwofn#1#2{{(#1 \subset #2)}}
\def\setsothersubsetothertwofn#1#2{{(#1 \subseteq #2)}}
\def\setsothermembertwofn#1#2{{(#1 \in #2)}}
\def\opohtwofn#1#2{{#1\circ#2}}
\def\opdividetwofn#1#2{{#1 / #2}}
\def\optimestwofn#1#2{{#1\times#2}}
\def\opdifferenceonefn#1{{-#1}}
\def\opdifferencetwofn#1#2{{#1-#2}}
\def\opplustwofn#1#2{{#1+#2}}

\def\optrianglerighttwofn#1#2{{#1 \strictif #2}}
\def\pvsid#1{\textrm{#1}}
\def\pvsdeclspacing{0in}		

\def\pvskey#1{\textbf{\uppercase{#1}}}

\begin{session}
  \(\optrianglerighttwofn{P}{Q}\): \pvsid{pmlformulas} \(\defn\) \(\neg\)\(\Diamond\)\pvsid{(}\(P\) \(\wedge\) \(\neg\)\(Q\)\pvsid{)}\vspace*{\pvsdeclspacing}

  \(\pvsid{H1}\): \pvskey{AXIOM} \(\optrianglerighttwofn{g}{\Box{}g}\)\vspace*{\pvsdeclspacing}

  \(\pvsid{H1\_triv}\): \pvskey{LEMMA} \pvsid{symmetric?}\pvsid{(}\pvsid{access}\pvsid{)} \(\supset\) \((\optrianglerighttwofn{P}{\Box{}P})\) \(\supset\) \((\Diamond\)\(P\) \(\supset\) \(P)\)\vspace*{\pvsdeclspacing}

  \(\pvsid{H2}\): \pvskey{AXIOM} \(\Diamond\)\(g\)\vspace*{\pvsdeclspacing}

  \pvsid{HC}: \pvskey{THEOREM} \pvsid{symmetric?}\pvsid{(}\pvsid{access}\pvsid{)} \(\supset\) \(g\)\vspace*{\pvsdeclspacing}
\end{session}

The way in which PVS ASCII text is rendered in \LaTeX\ is controlled
by a ``substitutions'' file \texttt{pvs-tex.sub}.  A comprehensive
substitution file is provided with the PVS distribution but it can be
augmented by the user.  The rendition above was generated using the
following substitutions file as augmentation.  The first column in
this file identifies the ASCII source to be substituted, the second
identifies the kind of PVS object it is (see the PVS documentation),
the third gives the size of the substitution in \emph{em}s, and the
final column gives the desired \LaTeX\ substitution.

\begin{sessionbiglab}{PVS \LaTeX\ substitution file}
\begin{verbatim}
|>               2      3       {#1 \strictif #2}
|>              id      1       \strictif
H1              id      2       \pvsid{H1}
H1_triv         id      3       \pvsid{H1\_triv}
H2              id      2       \pvsid{H2}
IMPLIES         id      3       \supset
=               key     2       \defn
~               id      1       \sim
~               id      1       \neg
=>              id      1       \supset
&               id      1       \wedge
\end{verbatim}
\end{sessionbiglab}

It might be considered that these substitutions are too aggressive
because they fail to distinguish those connectives that are to be
interpreted modally from those that are propositional.  If we remove
the last three lines from the substitution file above (where the
third-last line was overriding the fourth-last), we restore this
distinction and generate the following rendition.

\begin{session}
  \(\optrianglerighttwofn{P}{Q}\): \pvsid{pmlformulas} \(\defn\) \(\sim\)\(\Diamond\)\pvsid{(}\(P\) \(\&\) \(\sim\)\(Q\)\pvsid{)}\vspace*{\pvsdeclspacing}

  \(\pvsid{H1}\): \pvskey{AXIOM} \(\optrianglerighttwofn{g}{\Box{}g}\)\vspace*{\pvsdeclspacing}

  \(\pvsid{H1\_triv}\): \pvskey{LEMMA} \pvsid{symmetric?}\pvsid{(}\pvsid{access}\pvsid{)} \(\supset\) \((\optrianglerighttwofn{P}{\Box{}P})\) \(\supset\) \((\Diamond\)\(P\) \(\Rightarrow\) \(P)\)\vspace*{\pvsdeclspacing}

  \(\pvsid{H2}\): \pvskey{AXIOM} \(\Diamond\)\(g\)\vspace*{\pvsdeclspacing}

  \pvsid{HC}: \pvskey{THEOREM} \pvsid{symmetric?}\pvsid{(}\pvsid{access}\pvsid{)} \(\Rightarrow\) \(g\)\vspace*{\pvsdeclspacing}
\end{session}

We have now completed our treatment of propositional modal logic in
PVS and proceed to tackle the quantified case.

\section{Quantified Modal Logic in PVS}
\label{qml}

Quantified modal logics add quantification to the propositional case.
There can be interactions between quantifiers and the modal qualifiers
so this needs to be done with care.  For example, a standard step in
modal formulations of of Anselm's \emph{Proslogion} II argument for
the existence of God \cite{Eder&Ramharter15} (the ``Ontological
Argument,'' not to be confused with his Proslogion III argument which
we used for illustration in the previous section) is to consider
``some thing $x$ than which there is nothing greater.''  This might be
formulated as $\neg \exists y: \Diamond (y>x)$, which can be read as
``there is no $y$ that is greater than $x$ in any (accessible)
possible world.''  A plausible alternative is $\neg \Diamond \exists
y: (y>x)$, ``in no (accessible) possible world is there a $y$ greater
than $x$'' and we might wonder if this is equivalent to the previous
formula.  It turns out that sometimes it is and and sometimes it is
not, thereby highlighting the delicacy of combining quantified and
modal reasoning.

We saw that the full semantics for propositional modal logic must
recognize that possible worlds may not all be accessible from each
other, and this is formalized in the accessibility relation.  In a
similar manner, a precise semantics for quantified modal logic must
recognize that the domains of quantification may not be the same in
all possible worlds.  In a \emph{constant domains} semantics they are
the same, whereas in a \emph{varying domains} semantics they may
differ.  Important special cases are \emph{nondecreasing} and
\emph{nonincreasing} varying domains where the domains behave as their
names suggest across the accessibility relation (obviously, constant
domains are a special case of both of these).  The two formulas
considered above are equivalent if constant domains are assumed; with
varying domains, the first implies the second under nonincreasing
domains and vice-versa under nondecreasing domains.

Just as properties of the accessibility relation correspond to
standard modal formulas (e.g., symmetry corresponds to modal Axiom
B), so properties of domains correspond to standard formulas.   These
are the \textbf{Barcan Formula}
$$ \forall x: \Box \phi(x) \supset \Box \forall x: \phi(x)$$
or, equivalently, its contrapositive
$$\Diamond \exists x:\phi (x) \supset \exists x:\Diamond \phi (x)$$
and the \textbf{Converse Barcan Formula}
$$ \Box \forall x: \phi(x) \supset \forall x: \Box \phi(x)$$
and its contrapositive equivalent
$$ \exists x:\Diamond \phi (x) \supset \Diamond \exists x:\phi (x).$$
The Barcan formula characterizes nonincreasing domains, while the
Converse Barcan formula characterizes nondecreasing domains; the two
are equivalent if the accessibility relation is symmetric (i.e., if
modal Axiom B holds).

The Barcan formula and nonincreasing domains are generally regarded
with suspicion: they imply that anything that exists in a possible
world also exists in the actual world, a position known as
\emph{actualism}.  Thus, if it is possible that a cow jumped over the
moon, then there is a specific cow in the actual world that possibly
jumped over the moon.  The converse Barcan formula is less
controversial.

\subsection{Constant Domains}

To represent these topics in PVS, we begin with constant domain
semantics.  We take the full shallow embedding of Figure \ref{full-shallow},
rename \texttt{pmlformulas} to \texttt{qmlformulas}, and add the
following declarations.
\begin{sessionlab}{PVS fragment}
  QT: TYPE
  qmlpreds: TYPE = [QT -> qmlformulas]
  PP: VAR qmlpreds

  CFORALL(PP)(w): bool = FORALL (x:QT): PP(x)(w)
  CEXISTS(PP)(w): bool = EXISTS (x:QT): PP(x)(w)
\end{sessionlab}

Here, the type \texttt{QT} is the ``domain'' over which quantification
occurs.  In a finished formalization it will be best to specify this
as a parameter to the embedding theory.  In classical logic,
quantification applies to predicates over the quantified type; in
quantified modal logic, everything is lifted to operate over possible
worlds, so quantification applies to \texttt{qmlpreds}, which are
functions from the quantified type to predicates on worlds (i.e.,
\texttt{qmlformulas}).

We then specify the constant domain universal quantifier
\texttt{CFORALL} as a function that takes a lifted predicate of type
\texttt{qmlpreds} and delivers a lifted Boolean of type
\texttt{qmlformulas}; the definition of this function is quite
natural, it applies classical universal quantification to all values of
the lifted predicate in the world concerned.  The corresponding
existential quantifier is specified similarly as \texttt{CEXISTS}.

We can now attempt to employ these definitions by stating and proving
some properties of the Barcan formula.  In a constant domain, both the
Barcan formula and its converse are true, so we can prove that the two
sides of the formula are equal.  What we would like to write is the
following.
\begin{sessionlab}{Not yet PVS}
  Barcan: LEMMA CFORALL(□ PP) = □ CFORALL(PP)
\end{sessionlab}
Unfortunately, this is incorrect.  The modal qualifier $\Box$ applies
to \texttt{qmlformulas} and on the left we have given it \texttt{PP}
of type \texttt{qmlpreds}; what would work here is \texttt{□ PP(s)}
for some \texttt{s}.  However, \texttt{CFORALL} applies to
\texttt{qmlpreds} and \texttt{□ PP(s)} is of type
\texttt{qmlformulas}; what we need here is a function that takes an
\texttt{s} and returns \texttt{□ PP(s)}.  Such a function is
\texttt{LAMBDA s: □ PP(s)}, so a correct statement of the Barcan
formula is the following  (the right side of the equality was already
type-correct).
\begin{sessionlab}{PVS fragment}
  Barcan1: LEMMA
    CFORALL (LAMBDA (s:QT): □ PP(s)) = □ CFORALL(PP)
\end{sessionlab}
The formula is proved by the commands shown below;
\texttt{apply-extensionality} proves two functions are equal by
showing that their values are equal when applied to each element of
their domain.
\begin{sessionlab}{PVS proof}
(skosimp) (apply-extensionality :hide? t) (grind :polarity? t)
\end{sessionlab}

Although \texttt{Barcan1} is correct, it is a little difficult to read
and rather more difficult to write.  Fortunately, PVS has capabilities
that considerably ease these difficulties.  First is the
\texttt{K\_conversion}; this is defined in the PVS prelude as the K
combinator of Combinatory Logic (indeed, it might be less confusing if
it were named \texttt{K\_combinator}).  When used as a conversion it
automatically supplies the \texttt{LAMBDA} construction of
\texttt{Barcan1}.  Thus the following is a valid rendering of our
original formulation.
\begin{sessionlab}{PVS fragment}
  CONVERSION K_conversion

  Barcan: LEMMA CFORALL(□ PP) = □ CFORALL(PP)
\end{sessionlab}

The second PVS capability that is useful here is its ability to use
higher-order predicates as ``binders.''  The binder corresponding to a
higher-order predicate is its name with \texttt{!} appended, so the
binder corresponding to \texttt{CFORALL} is \texttt{CFORALL!}.
Thus another way to render the Barcan formula is the following.
\begin{sessionlab}{PVS fragment}
  Barcan2: LEMMA
    (CFORALL! (s:QT): □ PP(s)) = □ CFORALL! (s:QT): PP(s)
\end{sessionlab}
The right hand side could also be written as just
\texttt{□ CFORALL(PP)}, as it was previously.

These alternative forms are purely syntactic variations and are
interpreted the same as \texttt{Barcan1} internally.

Now, it might seem that instead of saying the two sides of the Barcan
formula are equal in constant domains, we could say each implies the
other, using the \texttt{IFF} connective as follows.
\begin{sessionlab}{PVS fragment}
  Barcan3: LEMMA CFORALL(□ PP) IFF □ CFORALL(PP)
\end{sessionlab}
However, if we ask PVS to display the expanded forms (i.e., with
conversions applied) of \texttt{Barcan} and \texttt{Barcan3}, using
the command \texttt{M-x ppe} we see they are different.
\begin{sessionlab}{PVS PPE}
  Barcan: LEMMA CFORALL(LAMBDA\,(s: QT): □PP(s)) = □CFORALL(PP)

  Barcan3: LEMMA
    valid(CFORALL(LAMBDA (s: QT): □PP(s))) IFF valid(□CFORALL(PP))
\end{sessionlab}
The reason is that all types have an equality operator, so
\texttt{Barcan} (once conversions are applied) is type-correct as it
stands.  The connective \texttt{IFF}, however, requires arguments of
Boolean type, so PVS searches for a conversion that will take the
\texttt{qmlformulas} we have provided and deliver Booleans: it finds
that \texttt{valid} fits the bill.

We can repeat this experiment using implication as in the following
two examples.  
\begin{sessionlab}{PVS fragment}
  Barcan4: LEMMA CFORALL(□ PP) IMPLIES □ CFORALL(PP)

  Barcan5: LEMMA CFORALL(□ PP) => □ CFORALL(PP)
\end{sessionlab}
The first uses the keyword \texttt{IMPLIES} and the second the
operator \texttt{=>}.  These are synonyms when given Boolean arguments,
but we have overloaded \texttt{=>} to take \texttt{qmlformulas} also.
Thus, when we display their expanded forms, we see they are different,
\begin{sessionlab}{PVS PPE}
  Barcan4: LEMMA
    valid(CFORALL(LAMBDA (s: QT): □PP(s))) IMPLIES valid(□CFORALL(PP))

  Barcan5: LEMMA valid(CFORALL(LAMBDA (s: QT): □PP(s)) => □CFORALL(PP))
\end{sessionlab}
These different formulas are all true (and provable) in constant
domains but have slightly different meanings.  This raises the
question: which is the correct interpretation of the Barcan formula?
Most texts on the topic are quasi-formal and do not clearly resolve
(or recognize) the different possible interpretations.  My belief is
that \texttt{Barcan5} is the correct reading.  It is worth repeating
in this context our earlier observation (page \pageref{nonded}) that
the deduction theorem (which would relate \texttt{Barcan4} and
\texttt{Barcan5}) is not valid in modal
logic.\footnote{\texttt{Barcan4} and \texttt{Barcan5} are equivalent
in constant domains, but not in varying domains.}

\subsection{Varying Domains}

So much for constant domains; the more general treatment of quantified
modal logic uses varying domains.  We can formalize this in PVS by
introducing a higher order predicate \texttt{vind(w)} (standing for
``values in domain'') that defines the set of members of the type
\texttt{QT} that are defined in world \texttt{w}.  This predicate is
best specified as a parameter to the theory.  We then require that the
quantifiers are restricted to just the defined values.  The obvious
way to do this is the following.
\begin{sessionlab}{PVS fragment}
  vind(w)(a): bool

  VFORALL(PP)(w): bool = FORALL (x:QT): vind(w)(x) IMPLIES PP(x)(w)
  VEXISTS(PP)(w): bool = EXISTS (x:QT): vind(w)(x) AND PP(x)(w)
\end{sessionlab}
However, PVS has predicate subtypes, so the following is a better way
to express this.
\begin{sessionlab}{PVS fragment}
  VFORALL(PP)(w): bool = FORALL (x:(vind(w))): PP(x)(w)
  VEXISTS(PP)(w): bool = EXISTS (x:(vind(w))): PP(x)(w)
\end{sessionlab}
Here, \texttt{vind(w)} is a predicate and a predicate in parentheses
denotes the corresponding subtype; hence, \texttt{x:(vind(w))}
indicates that variable \texttt{x} ranges over the subtype of values
satisfying \texttt{vind(w)}.

As we know, there are relationships between the Barcan formulas and
the way domains change across the accessibility relation.  We define
these as follows.
\begin{sessionlab}{PVS fragment}
  fixed: AXIOM vind(w)(a)

  nondecreasing: AXIOM (EXISTS w: vind(w)(a)\(\!\) AND\(\!\) access(w,v)) => vind(v)(a)

  nonincreasing: AXIOM (EXISTS w: vind(w)(a)\(\!\) AND\(\!\) access(v,w)) => vind(v)(a)
\end{sessionlab}
The axiom \texttt{fixed} asserts that these varying domains are
actually constant: every value is defined in every world.  The axiom
\texttt{nondecreasing} says that new values may be added as we move
along the accessibility relation, but none are lost; nonincreasing
says the opposite.

Incidentally, these two formulas would mean the same if written
without explicit quantification, as follows.
\begin{sessionlab}{PVS fragment}

  nondecreasing_alt: AXIOM vind(w)(a) AND access(w,v) => vind(v)(a)

  nonincreasing_alt: AXIOM vind(w)(a) AND access(v,w) => vind(v)(a)
\end{sessionlab}
Most readers will know this, but as this is a tutorial it is worth
spelling it out.  PVS closes formulas by universally quantifying all
free variables at the outermost level.  Here, we have implications
where variables \texttt{a} and \texttt{v} are referenced on both
sides, but \texttt{w} is referenced only on the left.  If we wish to
explicitly quantify \texttt{w} in its narrowest scope, we must use
an existential, not a universal, because there is an implicit negation
in the left side of an implication.  Readers who are unfamiliar with
this may wish to examine these formulas (and prove their equivalence)
in the PVS prover.

The Barcan formula is equivalent to nonincreasing domains and the
converse Barcan formula to nondecreasing.  In one direction, we can
simply state an instance of the Barcan formula and prove it by citing
the relevant axiom on domains.
\begin{sessionlab}{PVS fragment}
  vBarcan_eq: LEMMA VFORALL(□ PP) = □ VFORALL(PP)

  vBarcan: LEMMA VFORALL (□ PP) => □ VFORALL (PP)

  vCBarcan: LEMMA  □ VFORALL (PP) => VFORALL (□ PP)
\end{sessionlab}
These are all easily proved by citing the axiom concerned and then
using \texttt{grind} as usual.

More interesting is to prove the relationships in the other direction.
We do this below.  Notice that as the Barcan formula is now part of a
larger formula, we must be explicit about the scoping of the
quantification of \texttt{PP} (previously we could just allow PVS to
close the formula).  As always, we must state the relevant formula
(here, \texttt{nonincreasing}) rather than just use its name.
\begin{sessionlab}{PVS fragment}
  vBarcanx: LEMMA
    (FORALL PP: (VFORALL (□ PP) => □ VFORALL (PP))) IMPLIES
      (FORALL v,a: (EXISTS w: vind(w)(a) AND access(v,w)) => vind(v)(a))
\end{sessionlab}

A proof for this is shown below.  First, we use a variant of
\texttt{grind} that performs no quantifier instantiations.  Then we
supply a carefully crafted instantiation for \texttt{PP} that will
lead to a contradiction (we have nested \texttt{LAMBDA}s because the
predicate is higher-order), and then instantiate the other variables
in a way that makes the contradiction manifest.
\begin{sessionlab}{PVS proof}
(grind :if-match nil)
(inst - "LAMBDA (z:QT): LAMBDA (w:worlds): NOT(z=a!1 AND w=w!1)")
(inst -1 "v!1")
(ground)
(inst -1 "w!1")
(grind)
\end{sessionlab}

For the Converse Barcan formula, we state the result as follows (note
that we have changed the quantification for \texttt{w}, \texttt{v} and
\texttt{a} in illustration of the point made earlier about
quantification on the left side of implications).
\begin{sessionlab}{PVS fragment}
  vCBarcanx: LEMMA
    (FORALL PP: □ VFORALL (PP) => VFORALL (□ PP)) IMPLIES
      (FORALL w, v, a: vind(w)(a) AND access(w,v) => vind(v)(a))
\end{sessionlab}
And the proof is shown below.
\begin{sessionlab}{PVS proof}
(grind :if-match nil)
(inst - "LAMBDA (z:QT): LAMBDA (w:worlds): NOT(z=a!1 AND w=v!1)")
(inst -1 "w!1")
(ground)
(("1" (inst - "a!1") (grind)) ("2" (grind)))
\end{sessionlab}

Given these results, we can prove that the Barcan and Converse Barcan
are either both valid or both false when the accessibility relation is
symmetric.
\begin{sessionlab}{PVS fragment}
  bothB: LEMMA symmetric?(access) IMPLIES
    ((FORALL PP: VFORALL (□ PP) => □ VFORALL (PP)) IFF
       (FORALL PP: □ VFORALL (PP) => VFORALL (□ PP)))
\end{sessionlab}
We leave proof of this as an exercise; the basic approach is to first
prove a similar result about the \texttt{nonincreasing} and
\texttt{nondecreasing} formulas (PVS can prove this automatically),
then use \texttt{VBarcanx} and \texttt{VCBarcanx} to extend this to
the Barcan formulas (which requires about a dozen PVS proof steps).

Our treatment of modal logic in PVS is now adequately complete.  We
have an embedding of quantified modal logic and have seen stated and
proved some of the standard results concerning properties of the
accessibility relation and of constant and varying domains.  We have
seen how the capabilities of PVS allow modal formulas to be presented
in a fairly standard and attractive way.  To finish, we need to
discuss some of the pragmatics of using these embeddings, including
the notions of \emph{rigid} versus \emph{flexible} functions and
predicates, and the use of quantification in mixed classical and modal
contexts.

\subsection{Pragmatics of Quantified Modal Logic in PVS}
\label{erm2}

Eder and Ramharter \cite[page 2,819]{Eder&Ramharter15} quote Bertrand
Russell as follows:
\begin{quote}
I have heard a touchy owner of a yacht to whom a guest, on first
seeing it, remarked: ‘I thought your yacht was larger than it is’; and
the owner replied, ‘No, my yacht is not larger than it is’.
\end{quote}
The issue here is that of comparing the value of an attribute of an
object in one world with its value in another possible world.  Here,
the guest is comparing the size of the yacht in the actual world with
its size in the world of his imagination, while the owner is rooted in
the actual world.  This kind of comparison across worlds is a topic
that arises quite often and there is a method for dealing with it that
I will describe in this section, along with some other topics that
arise in the construction of more complex specifications involving
quantified modal logic.

At the beginning of this section, we saw a couple of quantified modal
formulas that purported to capture the idea of ``some thing $x$ than
which there is nothing greater.''  One of them was $\neg \exists y:
\Diamond (y>x)$, which can be read as ``there is no $y$ that is
greater than $x$ in any (accessible) possible world.''  The problem
with this formulation is that $y$ may be greater than $x$ in some
worlds, but its greatness in those worlds is exceeded by the greatness
of $x$ in the actual world.  So what we really want to compare is the
greatness of $x$ in this, the \emph{actual world}, against that of
potential rivals in other possible worlds.

Eder and Ramharter propose the following definition \cite[Section
4.2]{Eder&Ramharter15} for the predicate $G$ that recognizes maximally
great things under this new interpretation.  Here $\g(x)$ is the
``greatness'' of $x$ and $\succ$ is an ordering (actually an
uninterpreted predicate) on greatness.
\begin{description}
\item[Def M-God 3:] $G x : \leftrightarrow 
\exists z\, (z = \g(x) \wedge
  \neg \Diamond \exists y\, (\g(y) \succ z))$
\end{description}
The quantified variable $z$ is used to capture the greatness of $x$ in
\emph{this} world, so that it can be compared to that of some $y$ in
another possible world.

Before we look at the combination of modal and quantified
constructions here, it is worth taking note of the functions and
predicates involved.  The greatness of a thing is different in
different worlds.  It is what is called a \emph{flexible} function and
its lifted form will be $\g(x)(w)$, giving the greatness of $x$ in
world $w$.  But $\succ$ is a fixed or \emph{rigid} predicate: it does
not depend on the world (and so its lifted form is just
itself).\footnote{Varying domains allow the possibility that a
constant $c$ in the domain of quantification does or does not exist in
different worlds; a possible further complication is that it may exist
but denote different objects in different worlds.  Thus $\g(c)$ could
change from one world to another due to either or both the flexibility
of $\g$ or of $c$.  Fitting and Mendelsohn give details
\cite{Fitting&Mendelsohn}.  Flexible constants add complexity to the
embedding but seem to have little expressive value and we omit them
(if $c$ denotes $a$ in some worlds and $b$ in others, we can replace
it by $c_a$ and $c_b$; the former always denotes $a$ and exists in
worlds where $c$ exists and denotes $a$, and mutatis mutandis for
$c_b$).}  consider the quantification appearing in this definition.
The first quantifier is over greatness, whereas the second is over
things.  Our PVS embedding of quantified modal logic was defined in a
theory which took a domain of quantification as a parameter.  Here we
have two domains, so it looks as if we may need to extend the
embedding. or find a way to combine two instances of the embedding
theory.  The latter will entail the presence of two different
interpretations for \texttt{qmlformulas} and, depending how the
parameterization is done, several other types, too.  It is feasible to
resolve these issues by supplying almost all types used in the
\texttt{full\_shallow\_qml} theory as parameters, but it seems
appropriate to first look for alternative solutions.

The whole purpose of the quantification over $z$ is to provide a rigid
value that can be compared to flexible ones; that is, to provide a
nonmodal (i.e., classical) context for evaluation of modal
expressions.  So what we really want to write as the body of
\texttt{MGod3}, our PVS rendition of M-God 3, is something like the
following, where \texttt{z} is a classical quantification outside the
modal expression.  We use varying domains (i.e., \texttt{VEXISTS!})
for generality.
\begin{sessionlab}{Flawed PVS}
  MGod3(x): qmlformulas = 
    EXISTS z: (z=g(x) & \(\sim\) <> VEXISTS! y: (g(y) > z))
\end{sessionlab}
There are two problems here: one is that the type of the expression
quantified by $z$ is \texttt{qmlformulas} whereas classical
quantification requires a Boolean expression; the other is that
\texttt{MGod3(x)} is declared to be of type \texttt{qmlformulas},
whereas the quantified expression is delivering a Boolean.  We solve
both problems by recognizing that \texttt{MGod3(x)(w)} (i.e., the
evaluation of \texttt{MGod3(x)} in world \texttt{w}) is a Boolean and
so is the modal expression quantified by \texttt{z} when applied to
the world \texttt{w}.  Hence we arrive at the following formalization.

\begin{sessionlab}{PVS fragment}
modal_eandr: THEORY
BEGIN
  things: TYPE+

  x, y: VAR things
  IMPORTING full_shallow_qml[things]
  w, v: VAR worlds

  greatness: TYPE+
  a, b, z: VAR greatness

  g(x)(w): greatness

  >(a, b): bool

  MGod3(x)(w): bool = 
       EXISTS z: (z = g(x) & \(\sim\) <> VEXISTS! y: (g(y) > z))(w)
\end{sessionlab}

The theory defines \texttt{greatness} as an uninterpreted type;
\texttt{g(x)(w)} is a flexible function that gives the
\texttt{greatness} of \texttt{x} in world \texttt{w}, and \texttt{>}
is the rigid ordering over \texttt{greatness} (although, since it is
uninterpreted, nothing says it is really an ordering relation).  Then
we define \texttt{MGod3} in the manner described above.  

But now notice that the subexpressions \texttt{z = g(x)} and
\texttt{g(y) > z} in \texttt{MGod3} are not type-correct: \texttt{z}
is of type \texttt{greatness} but \texttt{g(x)} and \texttt{g(y)} are
functions from \texttt{worlds} to \texttt{greatness}.  The
subexpressions become correct when we lift them to functions on
worlds.  The \texttt{K\_conversion} of PVS does this automatically, so
the following is the prettyprint-expanded form of the definition
above.

\begin{sessionlab}{PVS PPE}
  MGod3(x)(w): bool =
      EXISTS z:
        ((LAMBDA (x1: worlds[U_beings]): z = g(x)(x1)) &
          \wig <>VEXISTS! y: (LAMBDA (s: worlds[U_beings]): g(y)(s) > z))(w)
\end{sessionlab}

Thus, although our PVS specification for \texttt{MGod3} is
syntactically quite similar to Eder and Ramharter's definition M-God
3, we see that quite a lot of machinery, and thought, are required to
create this similarity.

If we no longer require similarity to Eder and Ramharter's definition,
then some simpler alternatives become available, such as the following.
\begin{sessionlab}{PVS alternative}
  MGod3_alt(x)(w): bool =
      (\wig <> VEXISTS! y: LAMBDA (s: worlds): (g(y)(s) > g(x)(w)))(w)
\end{sessionlab}
Here, we exploit the fact that we have the world \texttt{w} available
and can therefore name the greatness of \texttt{x} in the current
world directly as \texttt{g(x)(w)}, thereby obviating the need for the
quantification over \texttt{z}.  On the other hand, we have to lift
the whole expression to a function on worlds using an explicit
\texttt{LAMBDA}.

This version of Anselm's Ontological Argument uses three premises that
Eder and Ramharter formulate as follows.
\begin{description}
\item[ExUnd:] $\exists x G x$,
\item[PossEx:] $\forall x \Diamond E! x$, and
\item[Greater 5:] $\forall x\forall y(\neg E! x \wedge \Diamond E! y \rightarrow
\exists z\, (z = \g(x) \wedge \Diamond ( \g(y) \succ z)))$,
\end{description}
where $E! x$ is a flexible predicate indicating that $x$ ``exists in
reality.''  
The conclusion is the following.
\begin{description}
\item[ERC:] $\exists x (G x \wedge E! x)$.
\end{description}
The first two premises and the conclusion are transcribed quite
directly into PVS as shown below.
\begin{sessionlab}{PVS fragment}
  ExUnd: AXIOM VEXISTS! x: MGod3(x)

  re?(x)(w): bool
  PossEx: AXIOM VFORALL! x: <> re?(x)

  ERC: THEOREM VEXISTS! x: MGod3(x) & re?(x)
\end{sessionlab}
Here, \texttt{re?} is a flexible predicate that corresponds to $E!$ in
Eder and Ramharter's notation.

\texttt{ExUnd} says that in all worlds, some greatest thing is in
``the understanding'' (i.e., in the domain of quantification);
\texttt{PossEx} says that everything exists in reality in some
accessible possible world; \texttt{ERC} says that in all worlds, there
is some greatest thing that exists in reality.

Notice that formalization of these premises in PVS forces attention on
the types.  Thus, the quantification in \texttt{ExUnd} is
\texttt{VEXISTS} (i.e., modal) rather than \texttt{EXISTS} (i.e.,
classical), which might not have been apparent in Eder and Ramharter's
notation, and similarly in \texttt{PossEx} and \texttt{ERC}.

Greater 5 says that if $x$ does not exists in reality in the actual
world and $y$ does exist in reality in some possible world, then the
greatness of $y$ in some possible world exceeds that of $x$ in the
actual world.  Greater 5 uses a similar construction to M-God 3, where
quantification over $z$ is used to record the greatness of $x$ in this
world for comparison against that of $y$ in some possible world.  When
formalizing Greater 5 in PVS, it seems sensible to use what we learned
with \texttt{MGod3} and apply the same technique.  That is, we use
\texttt{MGod3} as a pattern in defining a similar function \texttt{M3}
that specifies the expression on the right of the implication in
Greater 5 as follows.
\begin{sessionlab}{PVS possibility}
  M3(x, y)(w): bool = 
       EXISTS z: (z=g(x) & <> (g(y) > z))(w)
\end{sessionlab}
Then we can use \texttt{M3} in a straightforward specification of
Greater 5 such as the following.
\begin{sessionlab}{PVS possibility}
  Greater5: LEMMA
     VFORALL! x: VFORALL! y: (\(\sim\)re?(x) & <>re?(y) => M3(x, y))
\end{sessionlab}
But observe that \texttt{M3} could equivalently be defined as follows.
\begin{sessionlab}{PVS possibility}
  M3(x, y): qmlformulas = 
    LAMBDA w: EXISTS z: (z=g(x) & <> (g(y) > z))(w)
\end{sessionlab}

What we will do is employ the body of this definition directly in the
specification of \texttt{Greater5} given above, thereby removing the
need for \texttt{M3} (but having benefited from its consideration).
This leads to the following PVS specification.
\begin{sessionlab}{PVS fragment}
  Greater5: AXIOM 
     VFORALL! x: VFORALL! y: (\(\sim\)re?(x) & <>re?(y)
       => LAMBDA w: EXISTS z: ((z=g(x) & <> (g(y) > z))(w)))
\end{sessionlab}

The general technique employed in these examples is to use modal
constructions where we can, then ``lower'' them to classical logic
through application to a suitable world when we need to use classical
quantification, and to ``lift'' them back up again with a
\texttt{LAMBDA} when they are part of a larger modal construction.
Observe that this is performed automatically by the
\texttt{K\_conversion} for the subexpressions involving \texttt{z}, as
it was in \texttt{MGod3} (and it would also supply the \texttt{LAMBDA
w:} had we omitted it).  

Some may think it would be better if the conclusion ERC referred to
the actual world, rather than (implicitly) all possible words, and
likewise ExUnd.  It might seem that we could specify this alternative
version of ExUnd as follows.
\begin{sessionlab}{Flawed PVS}
  here: worlds

  ExUnd: AXIOM VEXISTS! x: MGod3(x)(here)
\end{sessionlab}
This is incorrect, however.  If we invite PVS to expand out the modal
constructs with the command \texttt{(grind :if-match nil :exclude
"MGod3")}, we obtain the following sequent from this version of
\texttt{ExUnd}.
\begin{sessionlab}{PVS sequent}
  |-------
\{1\}   EXISTS (x: (vind(w!1))): MGod3(x)(here)
\end{sessionlab}
We see that PVS has (correctly) required this expression to be true in
all worlds, whose Skolemized representative is \texttt{w!1}, and this
yields an unhelpful constraint on the type for \texttt{x}.  A correct
specification is the following (note the crucial parenthesization
around \texttt{VEXISTS!}),
which causes \texttt{x} to have the more useful type
\texttt{(vind(here))}.
\begin{sessionlab}{PVS fragment}
  ExUnd: AXIOM (VEXISTS! x: MGod3(x))(here)
\end{sessionlab}
Similar care is required with the adjustment to \texttt{ERC}.

That concludes our examination of this example.  Our interest in it
here is purely as an illustration of several tricky topics in the use
of quantified modal logic.  Those who are interested in the argument
itself, and its proof, are referred to \cite{Rushby:modalont19}.

\section{Conclusions}
\label{conc}

PVS is able to embed modal logic in a way the supports its concepts
and much of its standard notation while providing the benefits of
mechanized checking and automated reasoning.  Although the embedding
of modal logic is fairly straightforward, checking and automation
reveal that use of the quantified logic, in particular, requires care,
especially when combined with classical quantification.  Thus, readers
who have studied and, perhaps, experimented with the examples in the
previous subsection may be inclined to agree with Lewis, who writes as
follows.
\begin{quote}
``Philosophy abounds in troublesome modal arguments---endlessly
debated, perennially plausible, perennially suspect. The standards of
validity for modal reasoning have long been unclear; they become clear
only when we provide a semantic analysis of modal logic by reference
to possible worlds and to possible things therein.  Thus insofar as we
understand modal reasoning at all, we understand it as disguised
reasoning about possible beings.''  \cite[p.\ 175]{Lewis70}
\end{quote}

Lewis then presents a formal development written directly in terms of
possible worlds.  In my opinion this goes too far: there is value in
the use of modal concepts and notation.  However, formalizing modal
arguments in PVS can reveal subtleties and unsuspected complexities.
In particular, use of conversions and overloading in PVS sometimes
produces different interpretations for apparently similar formulas, as
in the different readings of the Barcan formula, and it highlights the
care needed with the formulas of Section \ref{erm2}.  In my view,
these should be seen as benefits of formalization and mechanization,
not drawbacks: the subtleties and complexities are real, and are
exposed by formalization.  But, because it can be hard to get the
details of modal specifications correct, I highly recommend inviting
PVS to expand them out so that (in a compromise with Lewis' position)
you can check that their rendition in terms of possible worlds
corresponds to your intent.

Benefits of the PVS mechanization of modal logic are illustrated by
some published examples concerning the modal ontological argument
\cite[Section 4]{Rushby21:ijpr}, where flawed criticism of a valid modal argument
is shown to be due to an erroneous formulation of \emph{modus tollens}
(recall Section \ref{stdax}), and a first order modal construction
encounters some of the difficulties outlined here in Section
\ref{erm2}.

The embedding technique used here is totally standard (see, for
example Wikipedia \cite{standard-translation}), but PVS's facilities
for overloading, conversions, and binders support it in a particularly
attractive way.  Embeddings in other verification systems are
described by Benzm\"{u}ller and Woltzenlogel Paleo
\cite{Benzmueller&Paleo15:Coq}.

\subsection*{Acknowledgments}

I am grateful to N.\ Shankar and Sam Owre, the principal developers of
PVS, for much help and advice on both PVS and logic.

\bibliographystyle{modplain}

\begin{thebibliography}{10}

\bibitem{standard-translation}
{\em Standard Translation (of modal logic into first-order logic)}.
\newblock \url{https://en.wikipedia.org/wiki/Standard_translation}.

\bibitem{Benzmueller&Paleo15:Coq}
Christoph Benzm{\"{u}}ller and Bruno~Woltzenlogel Paleo.
\newblock Interacting with modal logics in the {Coq} proof assistant.
\newblock In {\em Computer Science---Theory and Applications: 10th
  International Computer Science Symposium in {Russia}, {CSR} 2015}, Volume
  9139 of Springer-Verlag {\em Lecture Notes in Computer Science}, pages
  398--411, Springer-Verlag, Listvyanka, Russia, July 2015.

\bibitem{Boulton92:embedding}
Richard Boulton, Andrew Gordon, Michael Gordon, John Harrison, John Herbert,
  and John~Van Tassel.
\newblock Experience with embedding hardware description languages in {HOL}.
\newblock In {\em Theorem Provers in Circuit Design ({TPCD} '92)}, pages
  129--156, North Holland, Nijmegen, The Netherlands, 1992.

\bibitem{Eder&Ramharter15}
G\"{u}nther Eder and Esther Ramharter.
\newblock Formal reconstructions of {St.\ Anselm's Ontological Argument}.
\newblock {\em Synthese}, 192(9):2795--2825, October 2015.

\bibitem{Fitting&Mendelsohn}
Melvin Fitting and Richard~L. Mendelsohn.
\newblock {\em First-Order Modal Logic}, volume 277.
\newblock Springer Synthese Library, 1998.

\bibitem{Lewis70}
David Lewis.
\newblock Anselm and actuality.
\newblock {\em No\^{u}s}, 4(2):175--188, May 1970.

\bibitem{PVS:language}
S.~Owre, N.~Shankar, J.~M. Rushby, and D.~W.~J. Stringer-Calvert.
\newblock {\em PVS Language Reference}.
\newblock Computer Science Laboratory, SRI International, Menlo Park, CA,
  September 1999.

\bibitem{PVS}
{\em PVS home page}.
\newblock \url{http://pvs.csl.sri.com/}.

\bibitem{Rushby:modalont19}
John Rushby.
\newblock Mechanized analysis of modal reconstructions of {Anselm's}
  traditional {Ontological Argument}.
\newblock Technical report, Computer Science Laboratory, SRI International,
  Menlo Park, CA, January 2019.

\bibitem{Rushby21:ijpr}
John Rushby.
\newblock Mechanized analysis of {Anselm's} modal ontological argument.
\newblock {\em International Journal for Philosophy of Religion}, 89:135--152,
  April 2021.
\newblock First published online 4 August 2020.

\end{thebibliography}

\end{document}